\documentclass[preprint]{aastex701}
\shorttitle{Discovery of hot water vapor emission in LHA 115-S 18}
\shortauthors{Arias et al.}
\submitjournal{\apjl}

\begin{document}

\title{High-resolution near-IR spectroscopy of the B[e] supergiant LHA 115-S 18: discovery of hot water vapor emission}

\author[orcid= 0000-0002-4016-2501,sname='Arias']{María Laura Arias}
\altaffiliation{Member of the Carrera de Investigador Científico, CONICET, Argentina.}
\affiliation{Instituto de Astrof\'{\i}sica La Plata, CCT La Plata, CONICET-UNLP, Paseo del Bosque S/N, 1900, La Plata, Argentina}
\affiliation{Facultad de Ciencias Astron\'omicas y Geof\'{i}sicas, Universidad Nacional de La Plata, Paseo del Bosque S/N, 1900, La Plata, Argentina}
\email[show]{mlaura@fcaglp.unlp.edu.ar}  

\author[orcid= 0000-0001-8797-7209, sname='Torres']{Andrea Fabiana Torres} 
\altaffiliation{Member of the Carrera de Investigador Científico, CONICET, Argentina.}
\affiliation{Instituto de Astrof\'{\i}sica La Plata, CCT La Plata, CONICET-UNLP, Paseo del Bosque S/N, 1900, La Plata, Argentina}
\affiliation{Facultad de Ciencias Astron\'omicas y Geof\'{i}sicas, Universidad Nacional de La Plata, Paseo del Bosque S/N, 1900, La Plata, Argentina}
\email{atorres@fcaglp.unlp.edu.ar}

\author[orcid= 0000-0002-4502-6330, sname='Kraus']{Michaela Kraus} 
\affiliation{Astronomical Institute, Czech Academy of Sciences, Fri\v{c}ova 298, 251\,65 Ond\v{r}ejov, Czech Republic}
\email{ michaela.kraus@asu.cas.cz}

\author[orcid=  0000-0003-2160-7146, sname='Cidale']{Lydia Sonia Cidale} 
\altaffiliation{Member of the Carrera de Investigador Científico, CONICET, Argentina.}
\affiliation{Instituto de Astrof\'{\i}sica La Plata, CCT La Plata, CONICET-UNLP, Paseo del Bosque S/N, 1900, La Plata, Argentina}
\affiliation{Facultad de Ciencias Astron\'omicas y Geof\'{i}sicas, Universidad Nacional de La Plata, Paseo del Bosque S/N, 1900, La Plata, Argentina}
\email{lydia@fcaglp.unlp.edu.ar}

\begin{abstract}

The post-main-sequence evolution of massive stars involves phases of intense, often eruptive mass loss, including the B[e] supergiant phase. These hot stars are surrounded by cool, dense circumstellar disks that host complex chemistry, producing both molecules and dust. Understanding the mass-loss history of B[e] supergiants is essential for constraining stellar evolution models, particularly regarding their final stages. Near-infrared CO band emission serves as a key tracer of disk dynamics, typically arising from the inner edge of the molecular disk or ring. However, the oxygen-rich environments of these stars also favor the presence of other molecules which trace regions complementary to those probed by CO.

In this work, we present high-resolution near-infrared spectra of the Small Magellanic Cloud B[e] supergiant LHA 115-S 18. Our analysis reveals rotationally broadened CO emission consistent with a Keplerian molecular ring, alongside strong hydrogen wind features in both H and K bands and numerous metallic emission lines. Notably, we report the first detection of hot water vapor emission in a B[e] supergiant. This finding indicates the existence of extended cool and dense regions in a harsh environment. A radial velocity offset between molecular and Pfund line emission further supports a binary system, with the molecular gas potentially being circumbinary. The discovery of hot H${_2}$O around the B[e] supergiant star LHA 115-S 18 challenges classical models on evolution and chemistry of massive binary stars and provides critical insight into mass-loss processes and molecular enrichment of the ISM.
\end{abstract}



\section{Introduction} 

LHA 115-S 18 is a  highly peculiar luminous star located at the Small Magellanic Cloud (SMC). Despite being the subject of numerous studies over the past several decades, its true nature remains controversial. \citet{1989A&A...220..206Z}  classified it as a  B[e] supergiant (B[e]SG) and derived its fundamental parameters: $T_{\rm eff}$ = 25 000 K, log g = 3.0, $E(B -V)$ = 0.4, $L_*$ = 3.0-4.6 $\times$ 10$^{5}$ L$_\odot$, $R_*$ = 33-36 R$_{\odot}$, and a zero-age main-sequence (ZAMS) mass of $M$ $\approx$ 35-40 $M_{\odot}$.  
The star displays significant photometric and spectroscopic variability on different time scales, from days to years, resembling in some aspects an LBV in its quiescent phase \citep{1996ApJ...470..597M,2015salt.confE..55B}. Most notable are the Balmer lines that change from a pure-emission to a P Cygni profile and the line of He\,{\footnotesize II} $\lambda$4686 that changes from absence of the line to emission with the same strength as H$\beta$ \citep{1977IBVS.1304....1S, 1981A&A....95..191A,1987A&A...176...59S}. This latter fact, together with the identification of an X-ray counterpart source \citep{2013A&A...560A..10C,2014MNRAS.438.2005M} and the presence of Raman-scattered emission lines in the optical range \citep{2012MNRAS.427L..80T} suggest that LHA 115-S 18 is probably a binary system composed by a B-supergiant and a hot compact object.
 Additionally, molecular emission bands of  titanium monoxide (TiO) and carbon monoxide (CO) in the optical and near-infrared spectra of LHA 115-S 18, respectively, were reported \citep{1989A&A...220..206Z, 1996ApJ...470..597M, 2013A&A...558A..17O}  suggesting the presence of cool, dense molecular gas in the circumstellar environment \citep{1996ApJ...470..597M, 2010MNRAS.408L...6L}. CO and TiO bandhead emission also seem to be variable \citep{2023Galax..11...76K,2012MNRAS.427L..80T}. The star also shows strong infrared excess due to dust \citep{2010AJ....139.1993K}.

The disks of B[e]SGs  present appropriate physical conditions to facilitate efficient molecule formation and dust condensation. In such environments, where the number of oxygen atoms is greater than the number of carbon ones (O/C$>$1), the latter are locked in CO molecules. 
Due to its high dissociation temperature ($\sim$ 5000 K), the CO molecule can persist much closer to hot radiation sources than all other molecules, making it an effective tracer of the inner rim of circumstellar molecular rings \citep{2000A&A...362..158K}. In addition, excess oxygen atoms are available for the formation of other molecules or compounds, such as silicon monoxide (SiO), which has the second highest dissociation energy after that of CO, and 
TiO. The emission from the SiO band at $\sim$ 4 $\mu$m was discovered in four Galactic B[e]SGs \citep{2015ApJ...800L..20K}, while emission from TiO bands in the optical spectral range was reported for five Magellanic Cloud B[e]SGs \citep{1989A&A...220..206Z, 2016A&A...593A.112K, 2018A&A...612A.113T, 2012MNRAS.427L..80T}. Although these detections have not been the result of a systematic study, they offer a promising avenue for the search for other molecules that could be sensitive tracers of the physical properties of the B[e]SG disk regions prior to the dust condensation zone. A good candidate for a more complex O-bearing molecule is water (H$_2$O). Water vapor features have been reported from the environments of evolved late-type stars, such as the extreme C-rich AGB star IRC+10216 \citep{2001Natur.412..160M}, the oxygen-rich AGB star W Hydrae  \citep{1996A&A...315L.237N}, the red supergiant VY Canis Majoris \citep{1999ApJ...517L.147N} and the yellow hypergiant HD 269953 \citep{2022BAAA...63...65K}. Most of the water detections cited above are in the infrared region in the $K$ band or beyond. 

In this work, we report the first detection of hot water vapor in the circumstellar disk of the SMC supergiant LHA 115-S 18. We present high-resolution near-IR spectroscopic observations of the star and model in detail the H$_2$O emission spectrum, together with CO molecular emission and Pfund line emission observed in the K-band.

\section{Observations and data reduction}
LHA 115-S 18 was observed with  IGRINS (Immersion GRating INfrared Spectrometer) mounted on the Gemini South telescope, on October 3 2021 under the program GS-2021B-Q241.
The IGRINS spectrograph fully covers the H and K bands in the near-IR (1.45 $\mu$m - 2.45 $\mu$m) in a single exposure. The slit size is  0.34$''$ $\times$ 5$''$  and  the resolving power is R $\approx$ 45000 \citep{2010SPIE.7735E..1MY,2014SPIE.9147E..1DP,2016SPIE.9908E..0CM}. 
The observations were taken in one ABBA nodding sequence, with an exposure time of 650 s per offset position, obtaining a mean S/N ratio of 80. 
Data were  reduced using the IGRINS PipeLine Package (IGRINS PLP; \citet{2017zndo....845059L}) that performs sky subtraction, flat fielding, bad-pixel correction, aperture extraction, and wavelength calibration using sky OH emission lines. 
To eliminate telluric lines, each science spectrum was divided by a telluric standard A0V star spectrum (HIP1714) observed close in time and at a similar air mass than that of the science target. 
As the strongest telluric features were not completely removed during the data reduction procedure with the IGRINS PLP,  we reprocessed this step using the "telluric" task from IRAF\footnote{IRAF is distributed by the Community Science and Data Center at NSF NOIRLab, which is managed by the Association of Universities for Research in Astronomy (AURA) under a cooperative agreement with the U.S. National Science Foundation} package. Still, some telluric residuals remain mainly as weak absorption features throughout the entire spectral range.
A model spectra of Vega, provided by the IGRINS PLP, was  then multiplied to the divided spectra, to correct for the intrinsic lines of the standard star, that are mainly H lines. The spectra were corrected from heliocentric velocity.

\section{Results}

\subsection{H and K-band atomic spectrum} 

\begin{figure*}
\centering
\includegraphics[angle=-90,width=0.78\paperwidth] {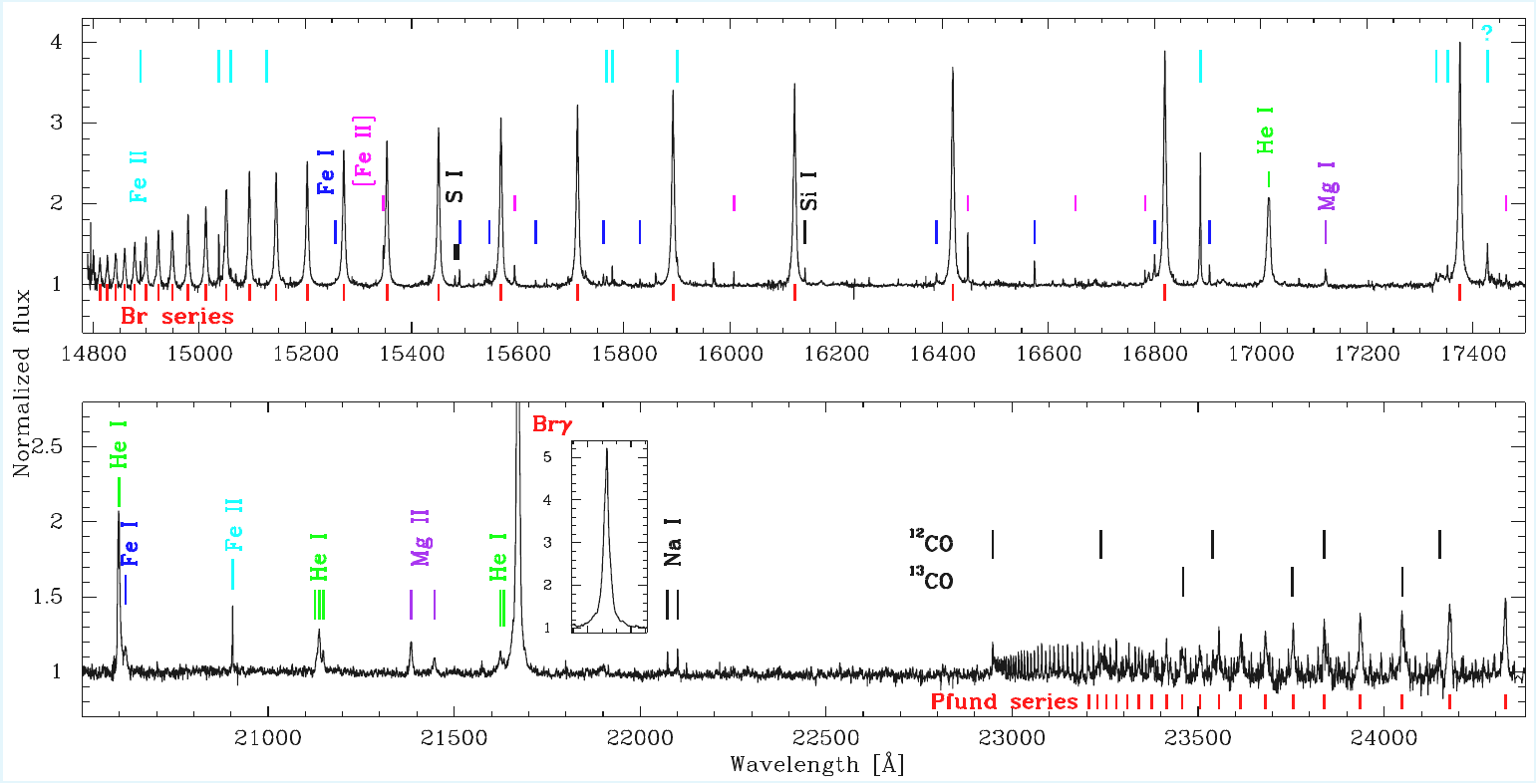}
\caption{GEMINI/IGRINS H-band (top) and  K-band (bottom) IGRINS high-resolution spectrum of the star LHA 115-S 18. Positions of the CO band heads, Pfund emission lines and  He\,{\footnotesize I} lines are marked with ticks and labeled. We also indicate the identified metallic emission lines.}
\label{S18-HK}
\end{figure*}

In Fig. \ref{S18-HK} we show  the  IGRINS high-resolution H- and K-band spectra of LHA 115-S 18, where we indicate the identification  of the strongest lines.
Pfund  (K-band) and Brackett (H-band) hydrogen lines appear in emission with one-peaked profiles. Also noticeable are He\,{\footnotesize I}  emission lines  at  1.700, 2.050, 2.112/113 and 2.161 $\mu$m, and several Fe\,{\footnotesize I}, Fe\,{\footnotesize II} and [Fe\,{\footnotesize II}] emission lines. The K-band spectrum shows emission lines from Mg\,{\footnotesize II} doublet at 2.138/144$\mu$m and Na\,{\footnotesize I} doublet at 2.206/209 $\mu$m, and a few other lines of neutral atoms. 
{These lines trace the atomic gaseous envelope revealing a stratified structure. The H\,{\footnotesize I} and He\,{\footnotesize I} lines and single ionized metallic lines forming in a highly ionized inner wind/disk, while Na I lines and other neutral atomic lines in an outer cooler region (T $<$ 5000 K) shielded from the star by the  dense inner layers.} 

\subsection{CO-band emission}
CO bands in emission are observed in the K-band spectrum of LHA~115-S~18. 
The high spectral resolution provided by IGRINS allowed us to resolve individual ro-vibrational lines of CO, in particular shortward of the second $^{12}$CO band head. Their profiles appear double-peaked (see Fig.~\ref{S18-fit}), suggesting rotation or equatorial in-/outflow of the gas.
Based on our experience with CO emission around evolved massive stars, where the line-forming region is usually confined in a narrow ring \citep[e.g.,][]{2012A&A...548A..72C, 2015AJ....149...13M, 2018A&A...612A.113T, 2021BAAA...62..104A}, we adopted the scenario of a rotating ring of molecular gas
for modeling the emission from LHA~115-S~18. 
We utilized the code developed by \citet{2000A&A...362..158K} for the computation of $^{12}$CO band emission  
from a circumstellar disk, modified by \citet{2009A&A...494..253K} and \citet{2013A&A...558A..17O} to add the emission of the isotopic molecule $^{13}$CO. 
The model considers that the CO gas is in local thermodynamic equilibrium, which is a suitable approximation for circumstellar environments, given their typical high column densities  \citep{2012MNRAS.426L..56O,2016A&A...593A.112K,2020MNRAS.493.4308K}. 
The K-band spectrum of LHA 115-S 18 also contains emission from the molecular isotope $^{13}$CO. The detection of measurable amounts of  $^{13}$C, locked into  $^{13}$CO  molecules, is an unambiguous tracer for chemically enriched material and age of the star \citep{2009A&A...494..253K,2010MNRAS.408L...6L}. We determined a $^{12}$CO/$^{13}$CO isotopic ratio of $5\pm1$. This value mirrors the stellar surface enrichment in  $^{12}$CO/$^{13}$CO at the time of mass ejection, indicating that LHA~115-S~18 is an evolved  (post-RSG) object.
Our best-fitting parameters to the CO emission are listed in Table~\ref{Table:param}.

\subsection{Emission from water vapor}

\begin{figure*}
\centering
{\includegraphics[angle=0,height=0.77\paperheight,width=0.82\paperwidth]{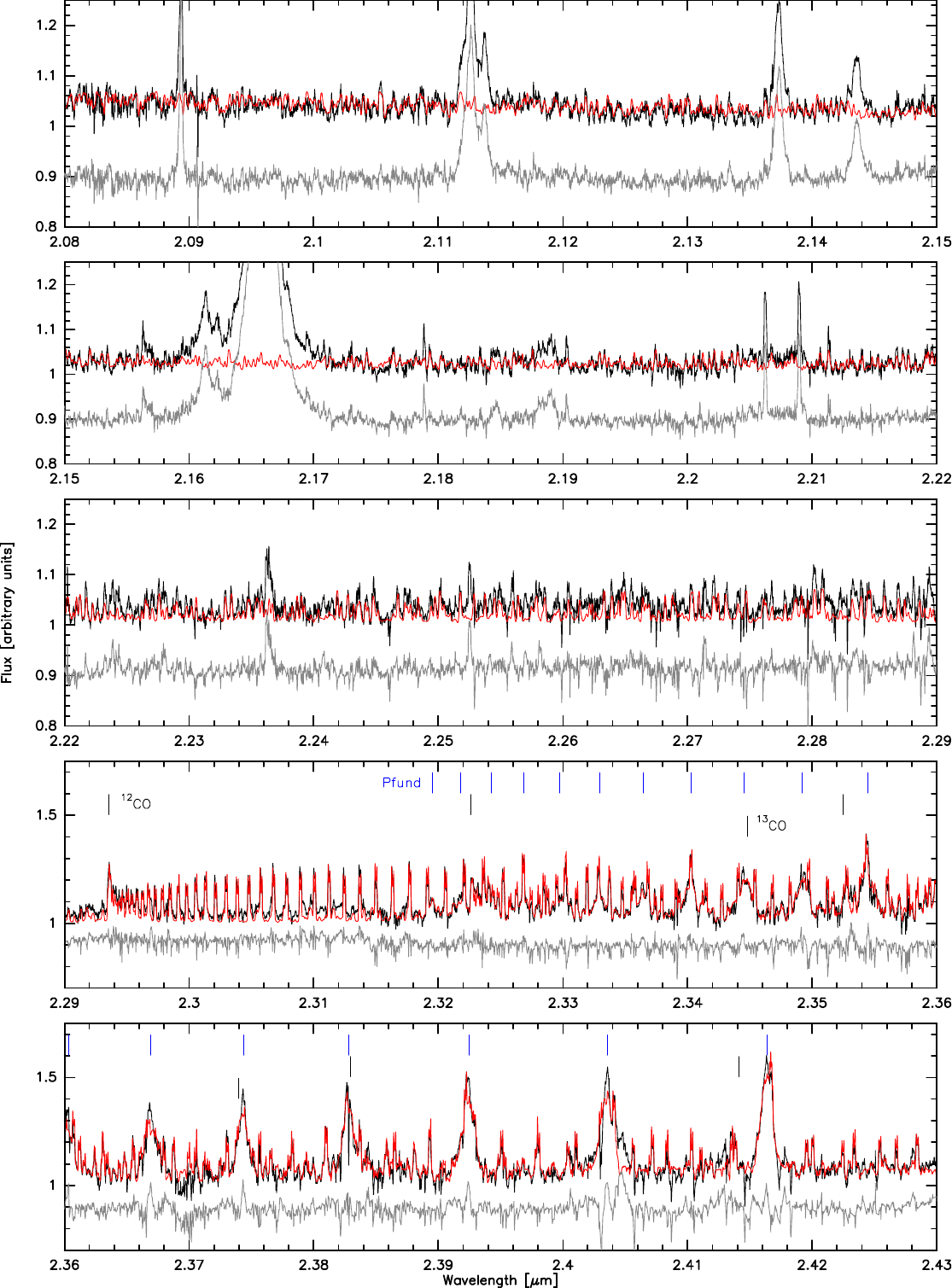} }
\caption{GEMINI/IGRINS observed K-band spectrum of LHA 115-S 18 (black), with the model composed of water
vapor (red, top three panels) and water vapor + CO bands + Pfund lines (red, bottom two panels) overplotted, along with the residuals
shifted for better visibility (gray). }
\label{S18-fit}
\end{figure*}

The K-band spectrum of LHA 115-S 18 also displays many small emission features shortward of and within the CO bands, which we identified as lines from water vapor (Fig. ~\ref{S18-fit}). We computed synthetic spectra for a model similar to that of  CO, i.e., considering a ring of gas with constant temperature and column density, and using the line list, energy levels, and Einstein transition coefficients from \citet{2018MNRAS.480.2597P}. 
{ This is the most complete line list of water spreading from the UV to the infrared that was computed in the frame of the ExoMol project\footnote{https://www.exomol.com }}. 
{ We have cropped this list to the wavelength
region of interest for our data, and it still    contains  more than 400 million lines that we have included in our computation of the emission
spectrum.} 

The water vapor lines show no indication for rotational broadening, so we use a Gaussian
profile with a velocity of $10\pm 0.5$ km s$^{-1}$, which might be interpreted as turbulent motion of the gas. The best-fitting parameters are listed in Table~\ref{Table:param}, and the synthetic 
spectrum of water vapor, which spreads over the full  observed wavelength range, is included in the synthetic spectrum shown in red in Fig.~\ref{S18-fit}, and is the sole fitting component in the top three panels. In these panels, most of the observed emission features can be identified with emission from water vapor. Also worth mentioning is the fact that the numerous water vapor lines all overlap, forming a quasi-continuum, so that basically no line-free continuum is evident throughout the entire spectrum.

\begin{figure*}
\centering
{\includegraphics[angle=0,width=0.78\paperwidth]{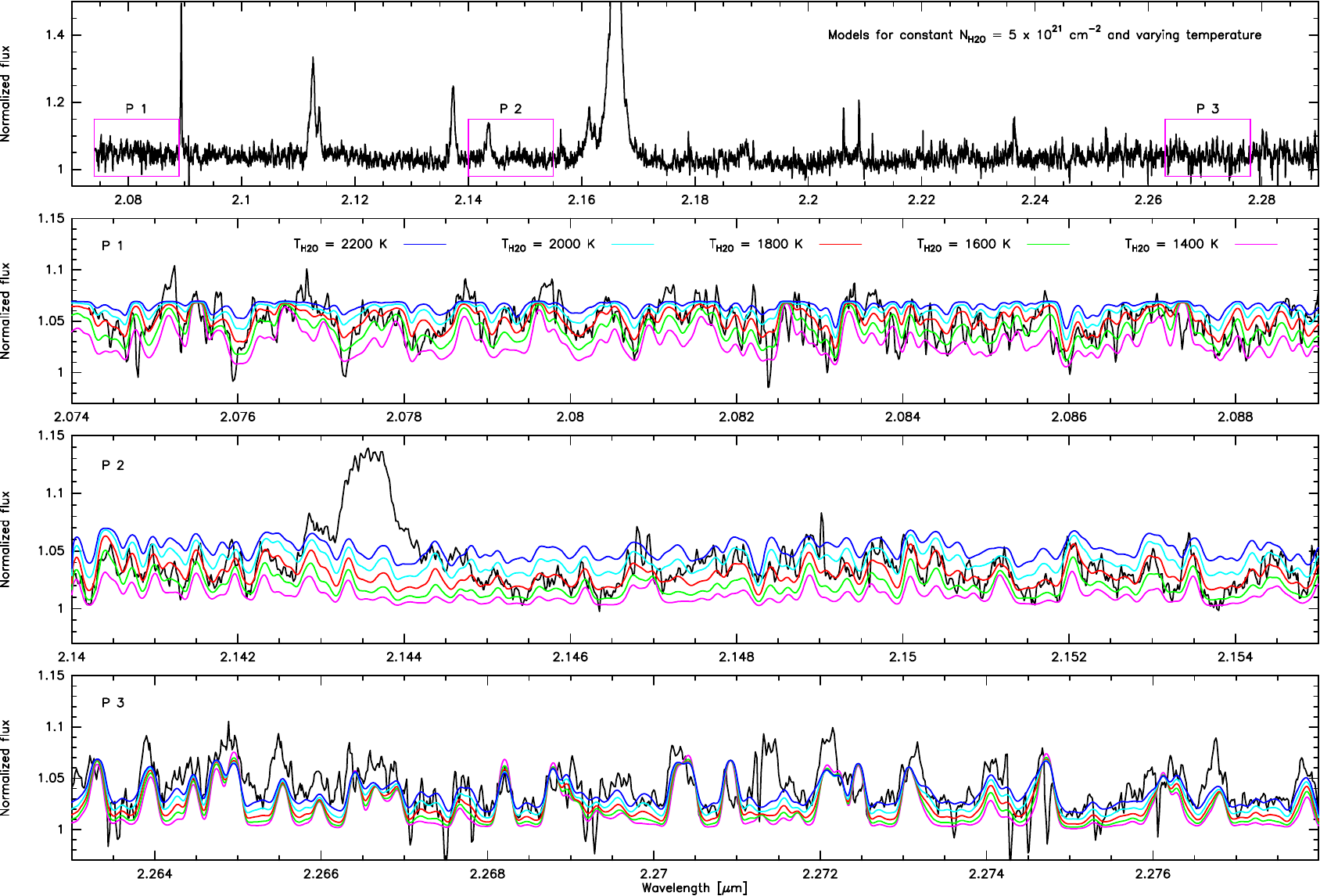} }
{\includegraphics[angle=0,width=0.78\paperwidth]{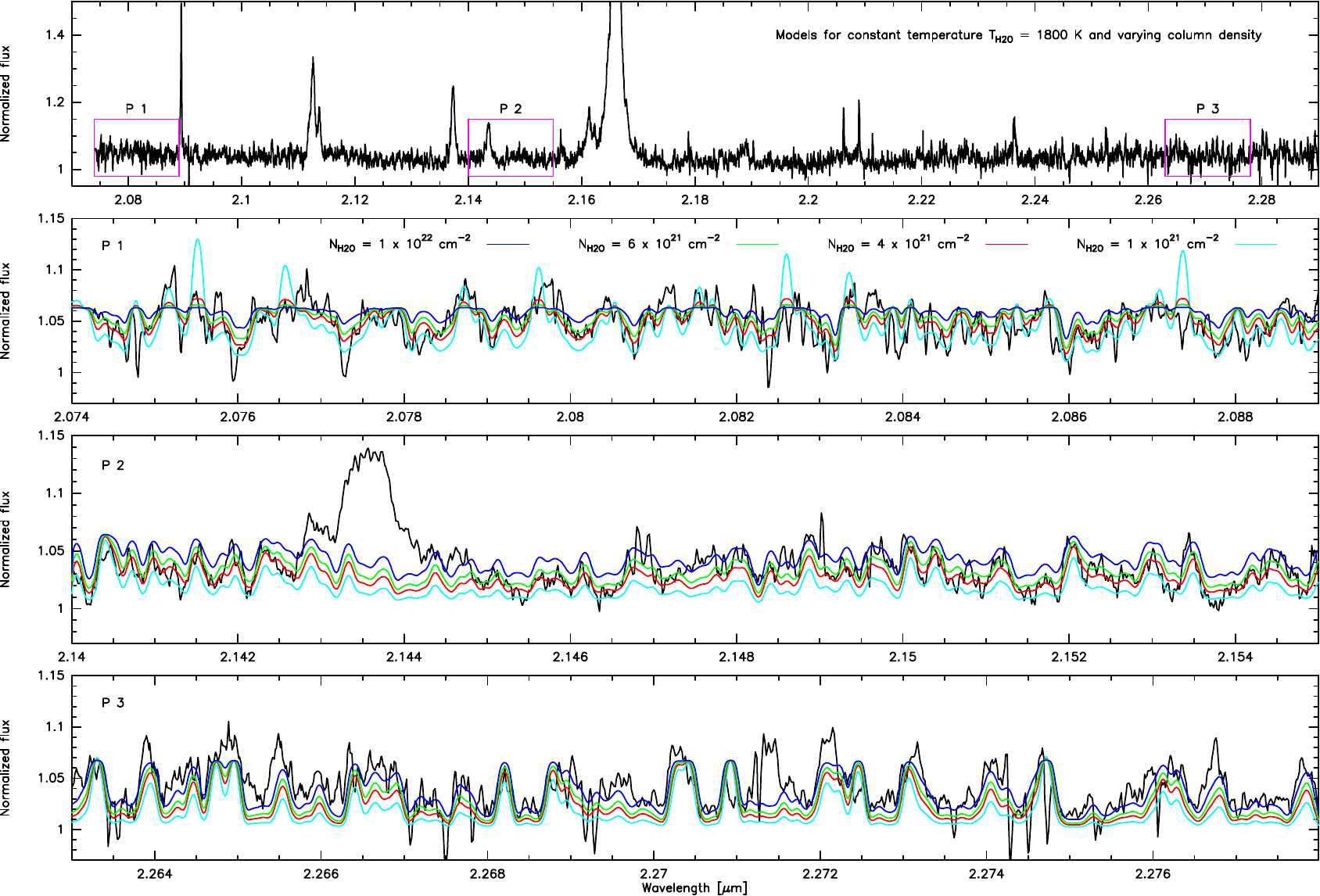} }
\caption{Different fits for selected portions of  LHA 115-S 18 K-band spectrum, showing the variation with the model parameteres: temperature and column density. The best fit is for T(H$_2$O) $\approx$ 1800 K and N(H$_2$O) $\approx$ $4 - 6 \times 10^{21}$ cm$^{-2}$.}
\label{fit-T-N}
\end{figure*}

{ To show the sensitivity of the model fits to the adopted parameters, we performed a
targeted exploration of the H$_2$O column density and temperature. To keep the computational effort manageable, this analysis was carried out for three representative wavelength intervals (P1, P2 and P3), which are indicated in the top panel of  Fig. \ref{fit-T-N}. 
For the temperature, we computed models for
T = 1400, 1600, 1800, 2000, and 2200 K while keeping the column density fixed at our best-fit value of N(H$_{2}$O) = $5 \times 10^{21}\rm{cm} ^{-2}$.

The comparison (shown in Fig. \ref{fit-T-N} four top panels) demonstrates that temperatures of 1400 K and 2200 K clearly fail to reproduce the observed emission pattern. The best agreement in regions P1 and P2 is obtained for T $\approx$ 1800 K, while models with $\pm$ 200 K already produce noticeable discrepancies in the relative intensity of the emission features. This supports our adopted uncertainty of approximately $\pm$200 K, which we consider a conservative estimate.

For the column density, we computed models for
N(H$_2$O) = $1 \times 10^{21}$, $ 4 \times 10^{21}$, $5 \times 10^{21}$, $6 \times 10^{21}$ and $1 \times 10^{22}$ cm$^{-2}$
at a fixed temperature of T = 1800 K. As illustrated in Fig. \ref{fit-T-N} (four bottom panels), the lowest column density corresponds to an almost optically thin case and significantly underpredicts the emission strength. In contrast, N(H$_2$O) = $1 \times 10^{22}$ cm$^{-2}$ approaches the optically thick limit in region P1. 
The best agreement is obtained for column densities around N(H$_2$O) $\approx$ $4 - 6 \times 10^{21}$ cm$^{-2}$, consistent with the value adopted in Table~\ref{Table:param}.}\\

\subsection{Pfund line emission}
The spectrum of LHA 115-S 18 displays emission from the hydrogen Pfund line series superimposed on the CO band spectrum. These lines show single-peaked profiles, suggesting that they form in a dense ionized wind rather than in a rotating or outflowing disk. 
This high-density environment can lead to pressure ionization effects, and hence to a sharp cut-off in the maximum number of detectable Pfund lines. Thus, the maximum detectable Pfund line is an indicator of the hydrogen density in the line-forming region.
To include the contribution of these lines to the total emission spectrum, we apply the code developed by \citet{2000A&A...362..158K} for the computation of the hydrogen series according to 
{Menzel case B recombination}, assuming that the lines are optically thin \citep{2010AJ....139.1993K}. The shape of the Pfund line emission spectrum is not sensitive to the electron temperature. We therefore fix the temperature at $T_e= 10\,000$~K, which is a typical value for an ionized wind. As the line profiles show no indication of rotational broadening, we adopt a pure Gaussian profile, which is a reasonably good approximation for optically thin recombination lines formed in a stellar wind. The maximum detectable Pfund line is Pf 38-5.  The parameters of the best-fitting model for the Pfund lines are listed in Table \ref{Table:param}.
We notice a blue-shift of the Pfund lines by about 25\,km\,s$^{-1}$ with respect to the molecular lines.

\begin{table}
\centering
\caption{Best-fitting model parameters.}
\begin{tabular}{lccc}
\hline\hline\noalign{\smallskip}
\!\!& \!\!\!\!CO&\!\!\!\!Water vapor\!\!\!\!&Pfund lines\\
\hline\noalign{\smallskip}
\!\!$T$ [K]  &  $2000\pm200$ & $1800\pm200$& -- \\
\!\!$N$ [$10^{21}$ cm$^{-2}$] & $0.2\pm0.05$& $5\pm 1$ &--\\
\!\!$v_\mathrm{rot,proj}$ [km\,s$^{-1}$] & $12\pm0.5$& --&--\\
\!\!$v_\mathrm{gauss}$ [km\,s$^{-1}$] & -- & $10\pm0.5$&--\\
\!\!     $^{12}$CO/$^{13}$CO& $5$ &--&--\\
\!\!$v_\mathrm{wind}$ [km\,s$^{-1}$] & -- & --&$80\pm0.5$\\
\!\!$T_e$ [K]  &  -- & --& 10$^{4}$\\
\!\!$N_e$ [$10^{12}$ cm$^{-3}$]  &  -- & --& 20\\
\hline
\end{tabular}
\label{Table:param}
\end{table}

\

\section{Discussion and Conclusions}

We present  high-resolution ($R \approx 45\,000$) simultaneous H- and K-band spectroscopic observations of the B[e] supergiant LHA~115-S~18, obtained using the IGRINS spectrograph.  Our data provide a detailed view of the near-infrared spectrum, dominated by intense wind emission lines from the Brackett and Pfund series and numerous emission lines from metals.
Moreover, the high quality of the spectroscopic observations of
LHA 115-S 18 allows us to clearly resolve rotationally broadened
emission from CO bandheads, consistent with a Keplerian rotating ring of
molecular gas, as is typical for this class of stars. By modeling this
CO emission, we were able to refine the physical parameters of its
emitting region.
The main result of this article is the identification of many weak emission features, shortward of and within the CO band emission region (from 2.45 to 2.75 $\mu$m), which correspond to hot water vapor.

While water vapor has been reported from the environments of evolved late-type stars, this is, to our knowledge, the first detection of water vapor from a B[e]SG star.
Modeling the water vapor emission, we found that the lower temperature and the narrower line profiles -when compared with those of the CO emission- indicate that the line-forming region lies farther away from the star. The measured shift in radial velocity between the molecular and the Pfund line emission seen in our spectrum supports a binary nature of the object. If this is the case, the molecular gas might be either circumstellar around the companion or circumbinary.

{ We also derive a column density ratio of N(H2O)/N(CO) $\approx$ 25 which is is broadly consistent with expectations for the oxygen-rich circumstellar environment  typical of B[e] supergiants. As, in these environments, C/O $<<$ 1 most of the available carbon is expected to be
locked into CO and the remaining oxygen can then participate in the formation of other oxygen-bearing molecules, including H$_2$O.
If the emitting area of the H$_2$O region is larger than
that of the CO region, the resulting column density ratio N(H2O)/N(CO) $\approx$ 25 is plausible.}
 
The discovery of water vapor emission in LHA~115-S~18 is so far unique among B[e] supergiants. Such an emission requires dense, heavily
shielded, cool (1000–3000 K) gas.
When considered together with the well-established presence of CO band emission, dust, and other molecular species, the detection of water vapor strongly suggests that the circumstellar/circumbinary disk of LHA 115-S 18 is chemically more akin to mass-loss products of cool evolved stars—such as yellow hypergiants (YHGs) or post–red supergiant objects—rather than to the outflows typically associated with classical B[e] supergiants. A compelling comparison is provided by the YHG HD 269953 in the Small Magellanic Cloud, which has recently been reported to exhibit hot water vapor emission in its K-band spectrum, with the  H$_{2}$O-emitting region located farther from the star than the CO-emitting zone, consistent with a rotating disk geometry \citep{2022BAAA...63...65K}. The observed $^{13}$CO enrichment in HD 269953 indicates that the circumstellar material is of stellar origin and was likely expelled during episodes of enhanced mass loss. It has also been suggested as a binary system \citep{2022MNRAS.511.4360K}.
LHA 115-S 18 exhibits a dense, strongly shielded molecular environment with an even lower $^{12}$CO/$^{13}$CO ratio, which may point to a related evolutionary history with a previous YHG phase.  However, LHA 115-S 18 is an exceptional representative of the B[e] supergiant class. In addition to the B[e] phenomenon, it displays large-amplitude spectroscopic and photometric variability, Raman-scattered line emission in tandem with strong variable He\,{\footnotesize II}  emission in the optical range, and weak but variable X-ray emission. Together, these properties indicate a highly complex circumstellar environment that is likely shaped, or at least strongly influenced, by binary interaction.
In this context, a combined scenario can be considered in which material ejected during a cooler evolutionary phase provides the molecular and dusty components of the circumstellar environment, while binary interaction—or a past common-envelope phase—may have contributed to the formation or long-term confinement of a dense, shielded disk. 
Thus, LHA 115-S 18 represents a key object for advancing our understanding of the late evolutionary stages of massive stars in low-metallicity environments.
Continued multi-wavelength monitoring of LHA 115-S 18 is therefore crucial for constraining its binary nature, the structure and dynamics of its circumstellar disk, and its potential evolutionary link between supergiant B[e] stars and other transitional massive-star phases. 

\facility{Gemini:South}
 
\begin{acknowledgments}
This work is based on observations obtained at the international Gemini Observatory, a program of NSF NOIRLab, which is managed by the Association of Universities for Research in Astronomy (AURA) under a cooperative agreement with the U.S. National Science Foundation on behalf of the Gemini Observatory partnership: the U.S. National Science Foundation (United States), National Research Council (Canada), Agencia Nacional de Investigaci\'{o}n y Desarrollo (Chile), Ministerio de Ciencia, Tecnolog\'{i}a e Innovaci\'{o}n (Argentina), Minist\'{e}rio da Ci\^{e}ncia, Tecnologia, Inova\c{c}\~{o}es e Comunica\c{c}\~{o}es (Brazil), and Korea Astronomy and Space Science Institute (Republic of Korea).
This work also used the Immersion Grating Infrared Spectrometer (IGRINS) that was developed under a collaboration between the University of Texas at Austin and the Korea Astronomy and Space Science Institute (KASI) with the financial support of the US National Science Foundation 27 under grants AST-1229522 and AST-1702267, of the University of Texas at Austin, and of the Korean GMT Project of KASI.
This project is co-funded by the European Union (Project 101183150 - OCEANS).
MLA, AFT and LSC
acknowledge the financial support from CONICET (PIP 1337) and the Universidad Nacional de La Plata (Programa de Incentivos 11/G192), Argentina, and MK from the Czech Science Foundation (GA \v{C}R, grant number 25-17532S). Thanks to the referee for his useful comments and suggestions.

\end{acknowledgments}

\bibliography{S18}{}
\bibliographystyle{aasjournalv7}



\end{document}